\documentstyle[preprint,prb,aps]{revtex}
\tightenlines
\begin{document}
\title{Surface critical behavior of two--dimensional dilute Ising
models}
\author{W. Selke}
\address{Institut f\"ur Theoretische Physik B, Technische Hochschule\\
D--52056 Aachen, Germany}
\author{F. Szalma, P. Lajk\'o}
\address{Institute for Theoretical Physics, Szeged University\\
H--6720 Szeged, Hungary}
\author{F. Igl\'oi}
\address{Research Institute for Solid State Physics, H-1525 Budapest,
P.O.Box 49, Hungary\\Institute for Theoretical Physics, Szeged University
H--6720 Szeged, Hungary}
 
\maketitle
 
\begin{abstract}
Ising models with nearest--neighbor ferromagnetic random couplings
on a square lattice with a (1,1) surface are studied, using 
Monte Carlo techniques and a star--triangle transformation method.
In particular, the critical exponent of the surface magnetization
is found to be close to that of the perfect model, $\beta_s$ =1/2.
The crossover from surface to bulk critical properties
is discussed.\\

\end{abstract}

\vspace{5mm}

\vspace{1cm}

\noindent {\bf 1. Model and methods}
 
The bulk critical behavior of the two--dimensional dilute Ising
model has been studied extensively in recent years. \cite{dot1,sha,se1,car}
According to renormalization group calculations, the randomness leads,
at least in the limit of weak dilution, to logarithmic modifications
of the asymptotic power--laws for various quantities in the perfect
model, in agreement
with results of Monte Carlo simulations (however, also conflicting
interpretations of numerical results have been
suggested and discussed \cite{se1,kuhn}). In
particular, the bulk magnetization, $m_b$, is expected to vanish as

\begin{equation}
m_b \propto t^{1/8} |ln t|^{-1/16}  
\end{equation}

\noindent
where $t$ is the reduced temperature, $t = (T_c- T)/T_c$.\\

In this Communication we shall present findings on surface critical
properties of nearest--neighbor random spin--1/2 Ising models
on a square lattice with a surface. Randomness is introduced
by allowing the nearest--neighbor ferromagnetic couplings to take
two values, $J_1$ and $J_2$, where $J_1$ is greater or equal to
$J_2$. If both couplings occur with the same probability, then 
the model is self--dual. \cite{frisch} The self--dual point is located at 

\begin{equation}
tanh (J_1/T_c) = exp (-2J_2/T_c)  
\end{equation}

\noindent
determining the critical temperature, if the model undergoes one
phase transition. Indeed, results of simulations \cite{wang} strongly
support that assumption.\\

Most of our findings are based on extensive Monte Carlo (MC) simulations, using
single--spin and cluster--flip algorithms. To facilitate comparison
of the simulational data with those of our numerical evaluation of
the star--triangle transformation (ST) method \cite{hill,ig1}, we
study the Ising model with a surface in the diagonal or (1,1) direction.
In that case, the coordination number of the surface spins is two.
(Indeed, we believe the critical properties at this ordinary surface
transition to be the same for the (1,1) and the (1,0) direction, as 
it is known to be the case in the perfect model).
In the simulations, we consider lattices consisting of $K$ columns and 
$L$ rows, where the first and last columns are surface lines; the first and 
last rows are connected by periodic boundary conditions. Usually, we
set $L = K/2$, with $K$ ranging from 40 to 1280. The ST method, which was
originally developed for layered lattices \cite{hill}, is generalized here
to treat general inhomogeneous systems. In these
calculations, $K$ is proportional to the number of iterations and
goes to infinity, while $L$, the number of surface sites,
remains finite. In both methods, MC and ST, one has to
average over an ensemble of
bond configurations. Typically, the number of realizations ranged,
in the simulations, from about 20 to several hundreds, taking more
configurations for smaller system sizes. In the single--spin flip
algorithm, used away from $T_c$, usually runs with a few $10^4$ Monte Carlo 
steps per site were performed. Closer to $T_c$, the more 
efficient one--cluster spin flip method was applied, taking into
account several $10^4$ clusters per realization. Note that 
the statistical errors during a MC run turned out to
be significantly smaller than those
resulting from the ensemble averaging. We tested different
random number generators to avoid inaccuracies due to a, possibly,
unfortunate choice of the generator. \cite{lan}\\

The crucial quantity, computed in the MC simulations, is the
magnetization per column, $m(i) = < |\sum s_{i,j}| >/L$,
where $s_{i,j}$ denotes the spin in column $i$ and row $j$,
with $i = 1, 2,... K$, and summing over 
$j = 1, 2, ... L$. Applying the ST method, we calculated, in particular, the
surface magnetization $m_s$ = $m(1)$. \\ 

In the following, we discuss the results of the MC simulations. The, so far
rather preliminary, findings obtained from the ST method are in  
very good agreement, demonstrating the correctness and accuracy of the
two approaches.\\

\noindent {\bf 2. Results}

The simulations were performed at $r = J_2/J_1$ = 1, 1/4, and
1/10, monitoring the effect of increasing randomness on the critical 
surface properties.\\

In all cases, the magnetization per column, $m(i)$, decreases as
one moves from the bulk towards the surface, as illustrated in
Fig. 1. For sufficiently wide systems, $K$, the magnetization
profile $m(i)$ displays a plateau around the center, $i = K/2$,
with the height being near the bulk magnetization, $m_b$. The bulk
magnetization is expected to be approached very closely at a distance $l$
from the surface, with the bulk correlation length
determining that distance. \cite{labi}\\

In the thermodynamic limit, on approach to the bulk critical
temperature, $T_c$, equation (2), $m(i)$ goes to zero. Close to $T_c$,
one may describe $m(i)$ by an effective
power--law behavior, $m(i) \propto t^{\beta(i)}$. Asymptotically,
for sufficiently small values of t, one has
$\beta(1) = \beta_s$, and $\beta(i) =  \beta$ for $i > l$ and
$i < K-l$. In general, one may define an effective, temperature
dependent critical exponent, $\beta(i)_{eff}$, by 

\begin{equation}
\beta(i)_{eff} = d ln [m(i)]/d ln [t]  
\end{equation}

\noindent
with the effective exponent becoming the asymptotic exponent as
$t$ vanishes. Certainly, in the MC study, $\beta(i)_{eff}$
can be only approximated from data at discrete temperatures,
say, $t$ and $t + \Delta t$. The resulting value of the effective
exponent is ascribed to the temperature $t + (\Delta t/2)$ (examples
are shown in Fig. 2). Since we are interested in the behavior in
the thermodynamic limit, the linear dimensions $K$ and $L$ have to
be sufficiently large compared to the bulk and surface correlation
lengths. Actually, to avoid finite size effects, we chose system sizes
with $L$ increasing approximately linearly with $1/t$ as one moves
towards $T_c$.\\

In the perfect case, $r =1$, our MC data for the magnetization $m(i)$
as well as the estimates for $\beta(1)_{eff}$, $\beta(2)_{eff}$, and
the effective exponent of the bulk magnetization,
$\beta_{eff}$, agree excellently with the exact
results \cite{mccoy,peschel}, see Fig. 3 (where we did not include
the simulational data), approaching smoothly, in
the limit of small t, the asymptotic exponents for the surface 
$\beta(1) = \beta_s = 1/2$ and the bulk $\beta =1/8$.
Note that $\beta(i)_{eff}$ decreases with $i$, at fixed
temperature, $t > 0$. There is an interesting
crossover phenomenon (which has not, to our knowledge, been studied
exactly, so far) in that effective exponent, being asymptotically
either 1/2 or 1/8, see Fig.2. The crossover occurs at a distance
from the surface reflecting the bulk correlation length
(that length diverges asymptotically
like $1/t$, i.e. in the same fashion as the surface correlation
length).\\ 

The effective exponents of the bulk and surface magnetizations, as
obtained from the simulations, for the
dilute cases $r = J_2/J_1$ = 1/4 and 1/10 are shown in Fig. 3. For
$r =1/4$, the values of the exponents, especially of $\beta(1)_{eff}$,
follow near criticality rather closely those
of the perfect model, $r =1$, as a function
of reduced temperature. However, perhaps most noticeably, the bulk
critical exponent $\beta_{eff}$ superceeds the asymptotic critical
value of the perfect case, $\beta = 1/8$, at $t < 0.05$, as had been
observed before. \cite{wang,talapov} Actually, the bulk magnetization
data coincide with the previous simulational results obtained for
the two--dimensional random--bond Ising model with full periodic
boundary conditions. Accordingly, $m_b$ can be fitted to the 
ansatz \cite{wang}

\begin{equation}
m_b = m_0 t^{1/8} (1+ at) ((1 +b ln[1/t])^{-1/16}  
\end{equation}

\noindent
with $m_0 = 1.203$, $a = -0.183$ and $b = 0.279$, where $b$
determines the crossover temperature to the critical region dominated
by randomness. On the other hand, the effective critical exponent of the 
surface magnetization $m_s$ continues to change gradually and 
smoothly, towards a value close to 1/2, as one enters that region.
An asymptotic exponent $\beta_s = 1/2$, as in the perfect case, seems
to be conceivable, without any logarithmic corrections to the simple
power--law.\\

At $r = 1/10$, i.e. at increased randomness, the effective exponent
$\beta(1)_{eff}$ tends to be almost constant over a wide range
of temperatures, $0.25 < t < 0.65$, being approximately 0.42, see
Fig. 3. Going closer to $T_c$, the exponents starts to increase
more visibly. Again, an asymptotic value of $\beta_s =1/2$ is
conceivable (certainly, strictly speaking, each extrapolation to
the limit of vanishing $t$, requiring exceedingly large lattice
sizes, is speculative). Note that $\beta_{eff}$ superceeds the
bulk exponent of the perfect case, 1/8, now already at $t < 0.12$.
Accordingly, we may safely argue to monitor, on further approach to $T_c$,
randomness--dominated critical behavior in the bulk and surface properties
as well. Obviously, in the case $J_2/J_1 = 1/10$, the results of the MC
simulations do not indicate that the critical exponent of $m_s$ is
strongly affected by randomness. A 'reasonable' guess
seems to be $\beta_s =0.49$, with an error bar of about 0.02.
Note that the error bars depicted in Fig. 3 are very pessimistic,
resulting from comparing the unfavorable $\sigma$--deviations of
the magnetizations at consecutive temperatures. Standard, more
optimistic, error analyses would reduce the size of the error
bars appreciably.\\

In summary, we conclude that our extensive MC simulations on
two--dimensional random--bond Ising models with a (1,1) surface
provide no compelling evidence for asymptotic critical exponents
depending strongly on the degree of dilution, i.e. the ratio of
the strength of the two different coupling constants. It seems
well conceivable that the surface magnetization follows the same
power--law behavior as in the perfect case, with $\beta_s = 1/2$,
without any logarithmic modifications. A detailed analysis, including
results from the star--triangle method for the surface magnetization
and the critical surface spin correlations, will be published elsewhere.\\ 
 
\noindent {\bf Acknowledgements}
  
P. L. and F. Sz. would like to thank the Institutes for Theoretical
Physics at the Universit\"at Hannover and at the Technische 
Hochschule Aachen for kind hospitality, and 
the DAAD and the Soros Foundation, Budapest,
for facilitating their visits. F.I.'s work was supported by the Hungarian
National Research Fund under grants OTKA TO12830 and OTKA TO23642.\\

\begin{figure}
\caption{Magnetization profiles for $J_2/J_1$ =1/4, at various
temperatures, $t$ = 0.05, 0.15, and 0.3, from bottom to top.
MC systems of sizes $L$ = 80 and $K$ = 160 were simulated.}
\label{fig1}
\end{figure}
 
\begin{figure}
\caption{Effective exponent of the magnetization per column,
for $J_2/J_1$ =1/4, at reduced critical temperatures $t$ = 0.275, 
0.175 and 0.075, from bottom to top. Systems with
$K$ =160 have been simulated.} 
\label{fig2}
\end{figure}
 
\begin{figure}
\caption{Effective exponents for the surface and bulk
magnetizations, for $J_2/J_1$ =1 (solid lines, exact results),
1/4 (full symbols), and 1/10 (open symbols). The dotted lines 
denote the asymptotic values of the perfect case. Systems with
$K$ = 80 (down triangles), 160 (up triangles), 320 (diamonds),
640 (circles), and 1280 (squares) have been simulated. The
error bars result from ensemble averaging.}
\label{fig3}
\end{figure}

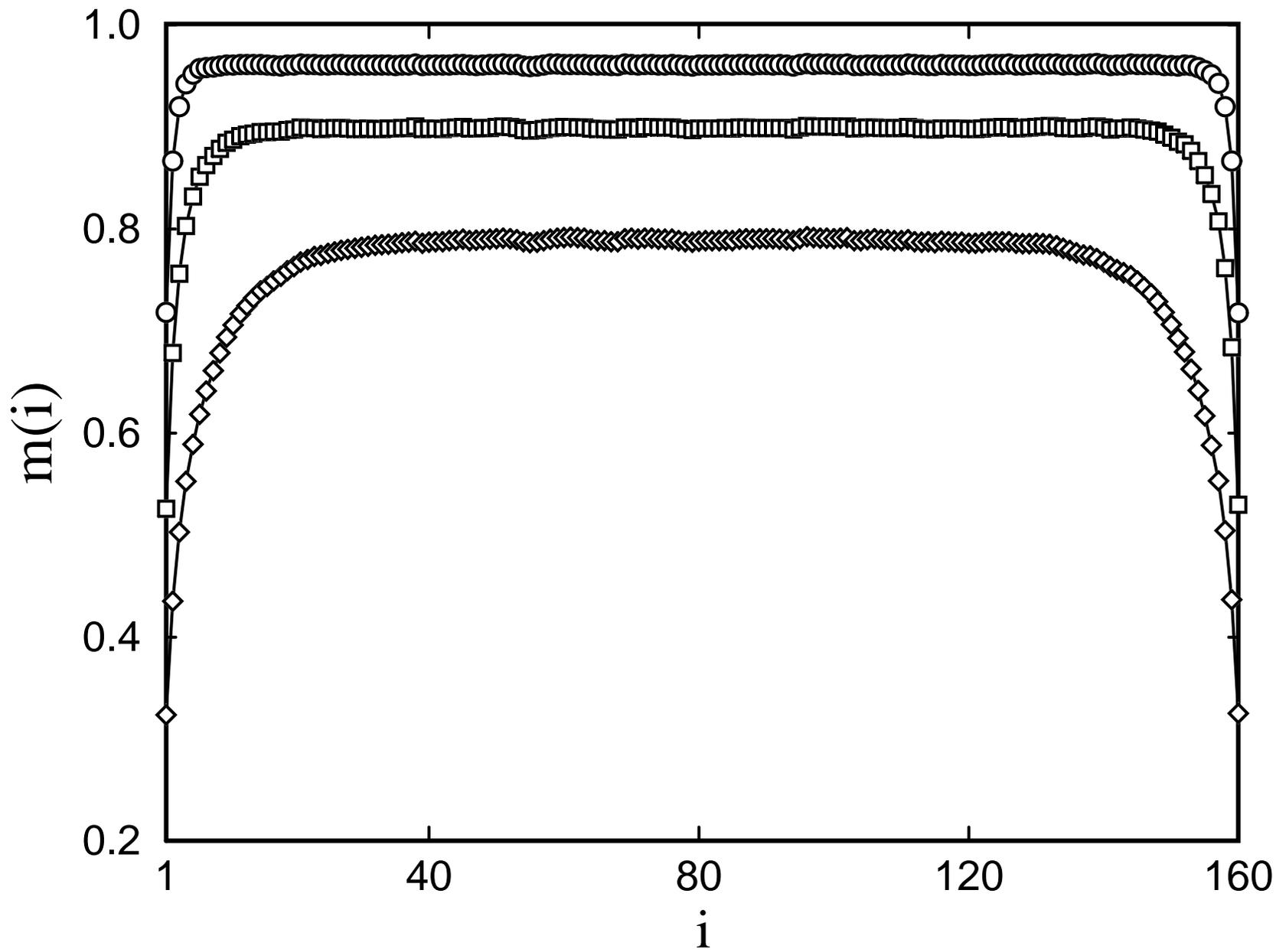
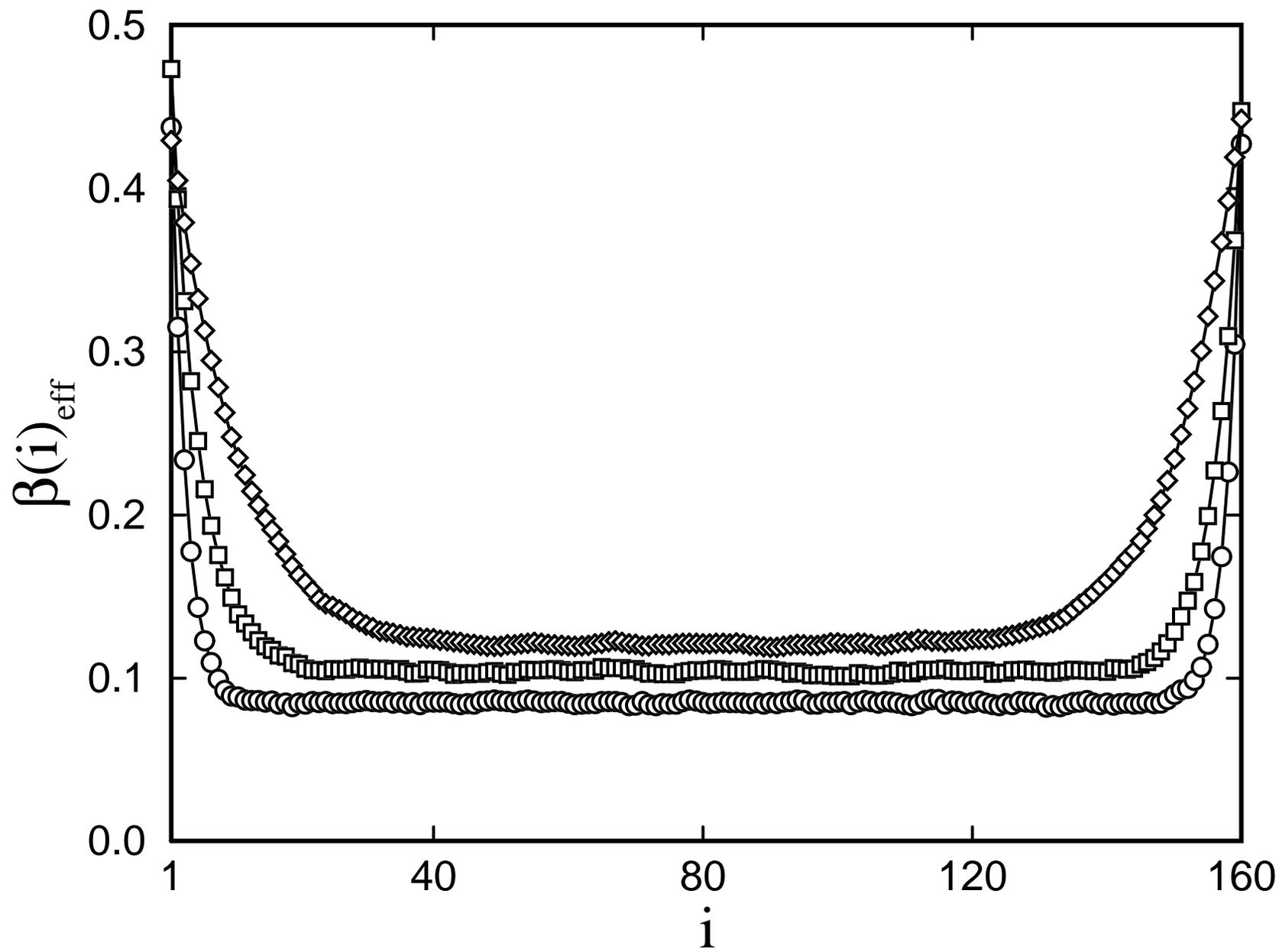
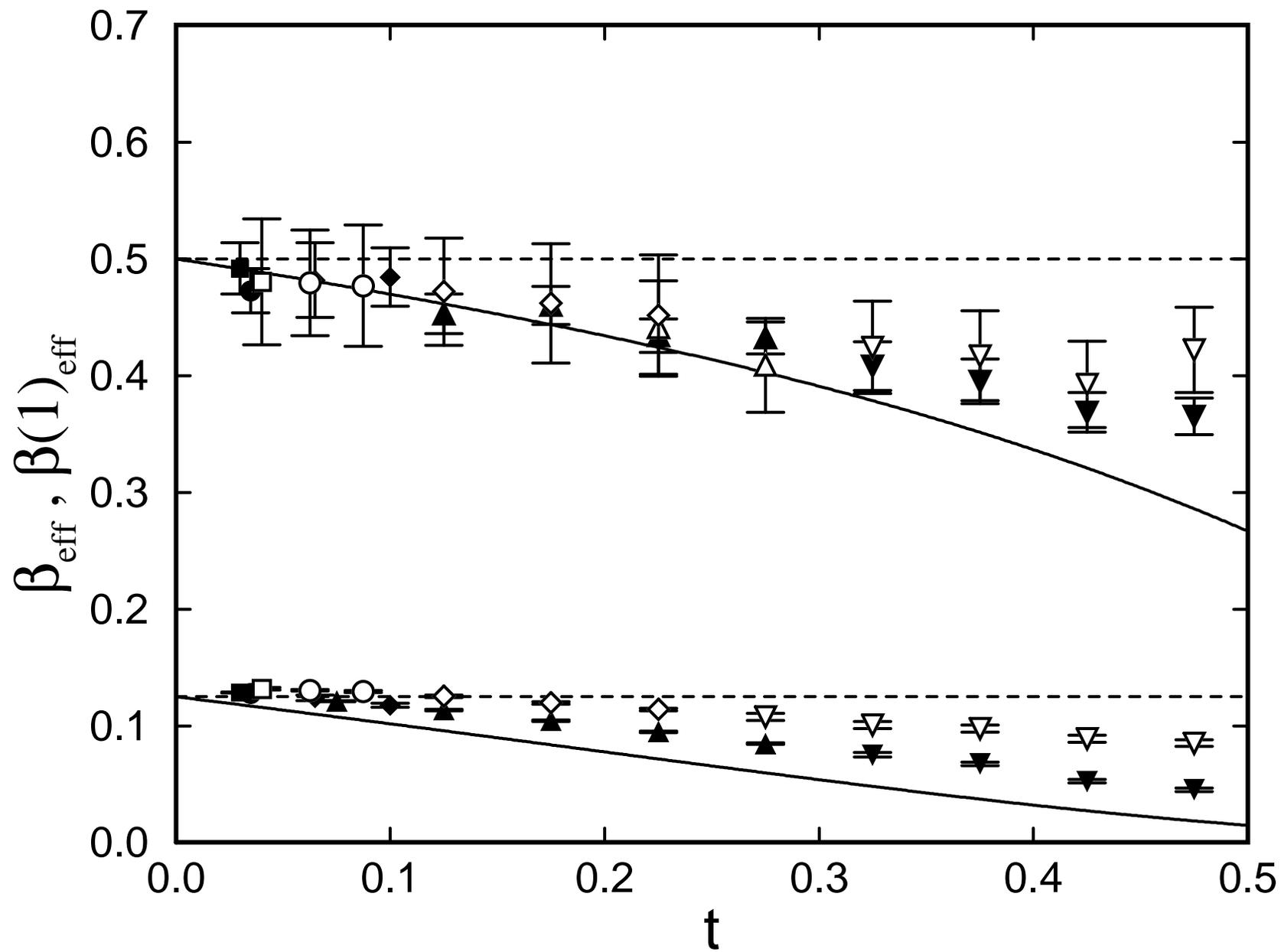
\end{document}